\documentclass[iop,apj]{emulateapj}
\usepackage{amsmath,amssymb,amstext}
\usepackage[markup=underlined]{changes}
\usepackage[breaklinks,colorlinks,citecolor=blue,linkcolor=magenta]{hyperref} 

\newcommand{\Msolh}{~h^{-1}{\rm  M}_\odot} 
\newcommand{\Msol}{~{\rm  M}_\odot} 
 
\newcommand{\Mpc}{~{\rm  Mpc}}

\usepackage[all]{hypcap} 

\usepackage{aas_macros}
\usepackage{natbib}
\shorttitle{Massive Cluster Strong Lensing}
\shortauthors{Li et al.}

\begin{document}

\title{The Importance of Secondary Halos for Strong Lensing in Massive Galaxy Clusters Across Redshift}

\author{Nan Li\altaffilmark{1,2,3,4}; 
Michael D. Gladders\altaffilmark{2,4}; 
Katrin Heitmann\altaffilmark{3}; 
Esteban M. Rangel\altaffilmark{7};
Hillary L. Child\altaffilmark{3,5} \\
Michael K. Florian\altaffilmark{6}; 
Lindsey E. Bleem\altaffilmark{3,4}; 
Salman Habib\altaffilmark{3}; 
Hal J. Finkel\altaffilmark{7}}

\altaffiltext{1}{School of Physics and Astronomy, University of Nottingham, University Park, Nottingham, NG7 2RD, UK; {\em \email{nan.li@nottingham.ac.uk}}}

\altaffiltext{2}{Department of Astronomy \& Astrophysics, The University of
Chicago, 5640 South Ellis Avenue, Chicago, IL 60637, USA;}
\altaffiltext{3}{High Energy Physics Division, Argonne National Laboratory, Lemont, IL 60439, USA}
\altaffiltext{4}{Kavli Institute for Cosmological Physics at the University of
Chicago, 5640 South Ellis Avenue, Chicago, IL 60637, USA}
\altaffiltext{5}{Department of Physics, The University of Chicago, 5640 South Ellis Avenue, Chicago, IL 60637, USA}
\altaffiltext{6}{Observational Cosmology Lab, NASA Goddard Space Flight Center, 8800 Greenbelt Rd., Greenbelt, MD 20771, USA}
\altaffiltext{7}{Argonne Leadership Computing Facility, Argonne National Laboratory, 9700 S. Cass Avenue, Lemont, IL 60439, USA}

\begin{abstract}
Cosmological cluster-scale strong gravitational lensing probes the mass distribution of the dense cores of massive dark matter halos and the structures along the line of sight from background sources to the observer. It is frequently assumed that the primary lens mass dominates the lensing, with the contribution of secondary masses along the line of sight being neglected. Secondary mass structures may, however, affect both the detectability of strong lensing in a given survey and modify the properties of the lensing that is detected. This paper focuses on the former: we utilize a large cosmological N-body simulation and a multiple lens plane (and many source plane) ray-tracing technique to quantify the influence of line of sight structures on the detectability of cluster-scale strong lensing in a cluster sample with a mass limit that encompasses current cluster catalogs from the South Pole Telescope. We extract both primary and secondary halos from the ``Outer Rim'' simulation and consider two strong lensing realizations -- one with only the primary halos included, and the other with the full mass light cone for each primary halo, including all secondary halos down to a mass limit more than an order of magnitude smaller than the smallest primary halos considered. In both cases, we use the same source information extracted from the Hubble Ultra Deep Field, and create realistic lensed images consistent with moderately deep ground-based imaging; the statistics of the observed strong lensing are extracted from these simulated images. The results demonstrate that down to the mass limit considered the total number of lenses is boosted by $\sim13-21\%$ when considering the complete multi-halo light-cone; the enhancement is insensitive to different length-to-width cuts applied to the lensed arcs. The increment in strong lens counts peaks at lens redshifts of $z\sim 0.6$ with no significant effect at $z<0.3$. The strongest trends are observed relative to the primary halo mass, with no significant effect in the most massive quintile of the halo sample, but increasingly boosting the observed lens counts toward small primary halo masses, with an enhancement greater than 50\% in the least massive quintile of the halo masses considered.

\end{abstract}
\keywords{keywords}

\maketitle
\section{Introduction}
In recent years, gravitational lensing has come to be a powerful tool to investigate the ``dark side'' of the Universe \cite[for reviews, see, e.g.,][and references therein]{Massey2010, Kneib2011, Treu2013, Meneghetti2013}.  Lensing effects can be observed over a wide range of scales: from mega-parsecs \cite[weak lensing,][]{Massey2007, Hoekstra2008, Okabe2010, Engelen2012, Mandelbaum2013, CFHT2013}, to kilo-parsecs \cite[strong lensing,][]{Treu2010,Suyu2010, Oguri2012, Coe2013, Newman2013, Kelly2015}, down to parsec scales \citep[micro lensing,][]{Muraki2011, Mao2012, Han2013, Gould2014}. It has been widely applied in extragalactic astrophysics and cosmology, e.g., in reconstructing the mass distributions of lenses \citep{Mandelbaum2006, Umetsu2008, Oguri2012, Newman2013, Han2015}, detecting galaxies at high redshift \citep{Richard2008, Jones2010, Jones2013, Stark2014}, measuring the Hubble constant \citep{Paraficz2010, Suyu2013, Suyu2014, Liao2015} as well as other cosmological parameters \citep{Mandelbaum2013, Fu2014, Jee2016, DESY12018}, amongst various applications.

Cosmological strong lensing is an extreme manifestation of this process, in which the mass density creating the potential -- i.e., the lens, which is typically a massive galaxy or a group or cluster of galaxies -- is sufficient to create multiple highly magnified and distorted images of background sources. The occurrence and morphological properties of these lensed images reflect both the properties of the gravitational potential between the source and the observer and the lensing geometry. It is a powerful probe of the central mass structure in galaxy clusters and groups and offers unique constraints on such systems \citep{Halkola2006, Sand2008, Newman2009, Limousin2010, Newman2011, Limousin2012, Bhattacharya2013, Grillo2015}.

Extended background sources -- i.e. galaxies -- typically produce arc-like images when strongly lensed, and the statistics and properties of such arcs are used in a variety of cosmological applications \citep{Kneib2011, Meneghetti2013}. The frequency of strongly lensed arcs on the sky reflects the abundance \citep{Dalal2004, Li2006, Fedeli2007, Hilbert2007, Fedeli2010}, the concentration \citep{Oguri2012, Sereno2013, Meneghetti2014} and astrophysical properties~\citep{Rozo2008, Peter2013} of massive lenses, and the redshift distribution and properties of the sources \citep{Wambsganss2004, Bayliss2011, Bayliss2012}. Thus, arc statistics help trace structure formation and can in principle provide constraints on cosmological parameters.

For efficiency and convenience it is often assumed that in the case of massive clusters the main lens dominates the lensing effects and single lens plane approximations are adopted to simplify the calculation of cluster-scale strong lensing \citep{Peirani2008, Horesh2011, Groener2014, Saez2016}. However, with the burgeoning sample of strong lenses now being discovered \citep[e.g.,][]{Diehl2017}, the contributions of structures along the line of sight to the strong-lensing cross section cannot be neglected when considering the precise statistics of large samples. Tools for the calculation of multiple-lens-plane ray-tracing \citep{Hilbert2007, Meneghetti2008, Giocoli2012, Petkova2014, McCully2014, Li2016, Plazas2019} have been developed to quantify such effects, but the calculated influence of line of sight structure spans a wide range \citep{Wambsganss2005,Hennawi2007, Hilbert2007,Faure2009, Puchwein2009, DAloisio2011, Jaroszynski2012, Jaroszynski2014, DAloisio2014, French2014, McCully2017, Birrer2017} likely due at least in part to the limited or specific sample of lenses considered. 

In this paper, we study the problem using an extremely large cosmological simulation -- the trillion-particle Outer Rim simulation -- that simultaneously combines a very large volume and the mass resolution necessary for strong lensing studies. We consider a primary halo sample down a mass limit consistent with the least massive galaxy clusters from the South Pole Telescope (SPT) \citep{Bleem2015}, i.e. down to a mass limit of 
$M_{500c}(\rho_{crit}) = 2.1\times 10^{14} \Msolh$, where $500c$ denotes the overdensity relative to the critical density $\rho_{crit}$, and $h = H_0/(100~km/s\Mpc^{-1})$. The influence of secondary halos along the line of sight on the total number of lenses, and the mass and redshift distributions of these lenses, is quantified by multiple lens plane ray-tracing simulations that consider all secondary halos in the light cone down to a mass limit of $M_{500c}(\rho_{crit}) = 7\times 10^{12}\Msolh$, more than an order of magnitude smaller than the least massive primary halos. 

This paper is structured as follows: in Section~\ref{sec:sims}, we describe the simulation and the strong lensing pipeline we are using; results are shown in Section~\ref{sec:results}, including the comparisons of the properties of simulated lensing systems with and without halos on the line of sight. A final discussion and set of conclusions can be found in Section~\ref{sec:sum}.   

\section{Simulations of Strong Gravitational Lensing}
\label{sec:sims}
Details of our strong lensing computational framework are presented in \cite{Li2016}. It consists of three main parts: cosmological simulations and the extraction of mass maps for the considered halos, a multi-lens-plane and many-source-plane ray-tracing pipeline, and an engine to create ``observed'' images that closely match real telescope data. In this section, we reintroduce the framework briefly, including a description of how we build the light cone of lenses and sources, calculate the deflection field, and implement the ray-tracing simulations through single lens planes and multiple lens planes.

\subsection{The Cosmological Simulation}
\label{sec:build_lc}

The cosmological simulation results used in this paper have been obtained with the Hardware/Hybrid Accelerated Cosmology Code \citep[HACC,][]{habib14, Habib2016}, a flexible, high-performance N-body code that runs on a range of supercomputing architectures. In this case, we used Mira, a BG/Q system at the Argonne Leadership Computing Facility to carry out the ``Outer Rim'' simulation, one of the largest cosmological simulations currently available.

The cosmology used is a $\Lambda$CDM model close to the best-fit model from WMAP-7~\citep{wmap7}. The cosmological parameters are: $\omega_{\rm cdm}=0.1109$, $\omega_{\rm b}=0.02258$, $n_s=0.963$, $h=0.71$, and $\sigma_8=0.8$. The comoving box size of the simulation is $L=4225.4~{\rm Mpc}=3000~h^{-1}{\rm Mpc}$, and it evolves 10,240$^3$=1.07 trillion particles, functioning as mass tracers. This leads to a particle mass of $m_p=2.6\times 10^{9} \Msol=1.85\times 10^9 \Msolh$. Extensive testing using a new tessellation-based density estimator \citep{Rangel2016} indicates that at this mass resolution we are able to reliably compute strong lensing for halos of masses $M_{500c}> 2\times 10^{14}\Msolh$. The simulation naturally incorporates substructure to much smaller mass scales; at these small scales the effects of (primarily stellar) baryons will play a role in the formation of small-scale lensing features. Future comparisons between the outputs of this pipeline and real data will consider such effects. The large volume of the Outer Rim simulation ensures that we have high-mass clusters at early times present in the simulation, and extensive statistics for massive systems that individually have large lensing cross sections. Unlike \cite{Meneghetti2008, Meneghetti2010} the baseline simulation used here is large enough to enable strong lensing calculations without resorting to re-simulation. 

Halos are identified with a Friends-of-Friends \citep[FOF]{Davis1985} halo finder with a linking length of $b=0.168$, versus the canonical value of $b=0.2$, following \cite{cohnwhite} who found that this reduced value mitigates the problem of halo overlinking. For halos with more than 100,000 particles, we save the complete information for all halo particles. Overdensity masses $M_{500c}$ of the clusters and secondary halos in the simulation are computed based on the centers of the FOF halos. These centers are determined by finding the potential minimum for each halo. We saved and analyzed 100 time snapshots between $z=10$ and $z=0$ evenly spaced in $\log_{10}(a)$ from the Outer Rim simulation. 

We employ 39 snapshots between $z=0$ and $z\sim 1.5$ in the analysis presented here. The fine sampling in time allows us to build halo light cones in post-processing. In order to enable building a full sphere we replicate the box eight times and place the observer at the center of the resulting cubes. The volume of the Outer Rim simulation allows us to build a light cone sphere in this way out to $z\sim 1.5$. While at high redshifts (beyond $z\sim 0.5$) a cluster appears more than once, the viewing angle is unique and therefore the projected cluster image and the intermediate mass between the observer and the cluster are different for each cluster investigated. In order to build the light cone, we read in each time slice separately. For each halo in a given time slice, we use its velocity to extrapolate the halo's position between the current and the next time step. If the halo's trajectory crosses the light cone we mark that position and store it. In this way, we build a light cone shell by shell until we reach the final timestep. One caveat with this method is that the trajectory of a given halo is not necessarily a straight line (if one were to follow its more detailed time evolution). Therefore a halo can occasionally appear twice on the light cone from two adjacent timesteps. Such instances have been identified and removed from the list of halos considered here. 

\subsection{Strong Lensing Image Pipeline}

The strong lensing simulation pipeline used here is PICS \citep[Pipeline for Images of Cosmological Strong lensing]{Li2016}. PICS can be utilized to generate realistic strong gravitational lensing signals from group- and cluster-scale lenses. The pipeline uses a low-noise, unbiased density estimator based on (resampled) Delaunay tessellations to calculate the density field \citep{Rangel2016}; deflection fields are estimated based on the surface density by using Equation 21-23 in \cite{Bartelmann2003}, and lensed images are produced by ray-tracing images of actual distant galaxies. 

The PICS pipeline requires source and lens light cones as inputs. For the image simulations presented here, source light cones are constructed by extracting galaxies from observed field of view. As in \cite{Li2016}, we utilize the Hubble Ultra Deep Field (HUDF) for a source population and images \citep{Beckwith2006}. Compared to other galaxy surveys with HST, e.g., The Cosmic Evolution Survey \citep[COSMOS,][]{COSMOS2007}, the HUDF is small, but provides excellent depth in multiple filters, and excellent photometric redshift grasp. Others have noted that the cosmic variance in HUDF is more significant compared to other possible HST deep fields (e.g., \citealt{Dobke2010} and \citealt{Moster2011}), however, the analyses which follow are purely comparative and insensitive to absolute source density. Note, we also do not attempt to de-convolve the HST PSF, or attempt to clean noise from the input images since the HST PSF is less than 1/10 the size of the PSF of the simulated ground-based images we generate, and the HUDF is many magnitudes deeper than those images.

To create a light cone of sources, we randomly select a position in the input images, and then extract a sub-light cone with $2048^2$ pixels, i.e., the angular size of the source light cone is about  $(61 \times 61)~ arcsec^2$. The galaxies in the light cone are divided into numerous source planes according to their redshifts from the catalog of \cite{Coe2006}, using the redshift grouping strategy detailed in \cite{Li2016}. On each source plane, constructed from a redshift bin in the source catalog, all galaxies have the same redshift -- namely the median redshift of all galaxies assigned to that plane. We set the number of source galaxies in each redshift bin to ten, which sets a reasonable balance between computational efficiency and bias. The image of each source is raytraced to the image plane, using the deflection field appropriate to the source plane to which it belongs. Any effects from a mismatch between the actual source redshift and the redshift interval to which the source is traced should be reduced to second order because the study we perform below is purely comparative between single and multi-halo light cones, and the same source population is used in both. Moreover, we note that the cross-section of gravitational lensing varies most rapidly at redshifts that also happen to have the highest galaxy counts, i.e., smaller redshifts bins at intermediate redshifts (where $N(z)$ is large) and larger bins at higher redshift (where $N(z)$ is small). Therefore, the bins of sources were selected using a simple calculation such that each bin has an equal number of galaxies. This allows for narrow redshift bins when lensing effects are rapidly changing, but also minimizes the number of redshifts for which deflection angle maps were required, though not precisely matched to lensing strength \citep{Plazas2019}.

We also create an aperture mask image for each galaxy in the simulation to aid in extracting lensed images of an individual source in the image plane. The aperture masks are created by connecting pixels at the source with a value more than $2\sigma$ above the background noise level, and keeping only pixels that have at least three neighboring pixels that are also at or above the threshold. These masks of the galaxies are raytraced to the image plane, labeled with unique integer values for each source.

A full sky light cone of all halos from the Outer Rim simulation is built as described in Section~\ref{sec:build_lc}. Based on this full sky light cone, we create individual lens light cones positioned on halo centers with masses $M_{500c}$ above $2.1 \times 10^{14} \Msolh$. This limit includes the least massive clusters in the SPT-SZ cluster samples \citep{Bleem2015}, though it is not an attempt to model the exact mass limit with the redshift of that sample (or incompleteness at that limit); we defer such work to a future paper. There are 10294 halos in the simulation light cone satisfying this cut. This is much larger than extant cluster samples at these masses \citep[e.g.,][]{Bleem2015}, sufficiently large that statistical uncertainties in the halo counts in the analyses presented here will be small by comparison to actual cluster samples. Unlike some previous work, where simulation statistics are boosted by considering a smaller number of massive halos over a range of projections, or by building a light cone by repeating (including rotating) the smaller boxes of simulation snapshots several times \citep{Hennawi2007, Hilbert2007, Puchwein2009, French2014}, each primary halo in the simulated sample considered here is raytraced only once, in the nominal lightcone in which it is found.

The size of the individual fields-of-view is $183\times183~arcsec$. For the case of a single lens plane,  we keep only the primary halo in each individual lens light cone. Lensed images and the corresponding lensed masks are produced by running single-lens plane ray-tracing through one lens plane from multiple source planes. For the case of multiple lens planes,  we include all halos more massive than $M_{500c} = 7\times10^{12} \Msolh$ into the lens light cone and place each lens on an individual lens plane. We exclude from consideration rare instances in which there are two halos along the same line of sight that both would be considered primary halos, or which jointly might produce a projected mass signal that would place them collectively in the primary halo list. Treating such instances fully requires a detailed treatment of the initial cluster selection process in the survey observations. However, at the sky surface densities for clusters of this mass range (one per few square degrees), such projections are rare, even when considering the spatial correlation of halos. Each source plane is ray-traced through all lens planes between the source plane and observer, creating lensed images as well as the corresponding lensed source mask. The PICS density estimator is sufficiently reliable for accurate strong lens ray-tracing when the number of particles in a halo is more than $\sim10^5$. At the Outer Rim particle mass of $1.85\times10^9 \Msolh$, the number of particles in halos with a mass above our chosen mass cut ($M_{500c} = 2.1 \times 10^{14} \Msolh$) is always sufficient for detailed ray-tracing \citep{Rangel2016}, but this is not the case for less massive halos. In both the single- and multi-halo cases, the surface density maps of primary lenses are modeled with particles. However, in the multi-lens plane case, the less massive halos along the line of sight are modeled with spherical NFW profiles \citep{Navarro1997} using the masses and concentrations measured from the particle distributions that correspond to each halo \citep{Child2018}.

The parameters of the output images are the same in both the single- and multi-halo raytrace cases. The pixel size of these initial lensed images is 0.09", and there are $2048^2$ pixels in the image plane, identical to the angular size of the individual lens light cones. The typical Einstein radius of these massive clusters is at most several tens of arcseconds; the output image size is sufficient to cover the strong lensing region of the primary halos, and to include the lensing effects of the line of sight environment. Following the procedure outlined above, lensed images are produced from the F435W, F606W, and F775W images of the input deep field data and these are mapped to the blue, green and red channels of the color images constructed below. 

\subsection{Arc Visibility in Visual Searches}
The visibility of strongly-lensed features is a function of observational parameters, primarily image depth and seeing. To extract a result from the simulated image set above, we thus choose particular values, and ``observe'' the set of potential strong lenses by convolving with an appropriate point spread function (PSF), pixelating to an appropriate scale, and adding noise to simulate a particular depth. We use a simple Gaussian PSF with a full width at half-maximum of 0.7" and a pixel scale of 0.2", typical for imaging instruments on ground-based telescopes at excellent sites. The depths are chosen to match that expected for 3-minute integrations using the Parallel Imager for Southern Cosmology Observations \citep[PISCO;][]{Stalder2014} on the Magellan telescopes, motivated by an ongoing program to image the entire SPT-SZ cluster sample \citep{Bleem2015} using that system. Note however that the analysis presented below is purely comparative -- i.e., the abundance and character of observed strong-lensing clusters with and without the effects of additional mass in the full light cone -- and that such effects are not expected to be strongly dependent on imaging depth or seeing. Thus, an exact match to a given survey or dataset is not required in this case. In particular, the limits chosen above are within factors of a few of those appropriate for strong lensing in the Dark Energy Survey \citep[e.g.,][]{Diehl2017} and so should be germane to strong lensing in that dataset as well. Furthermore, unlike the image examples that are shown in \cite{Li2016}, we do not in this instance add other (unlensed) galaxy images in the lightcone, or foreground stars, or other image-level effects. Again, this is because the analysis pursued here is purely comparative. 

\begin{figure*}
\centering
\includegraphics[width=0.98\textwidth]{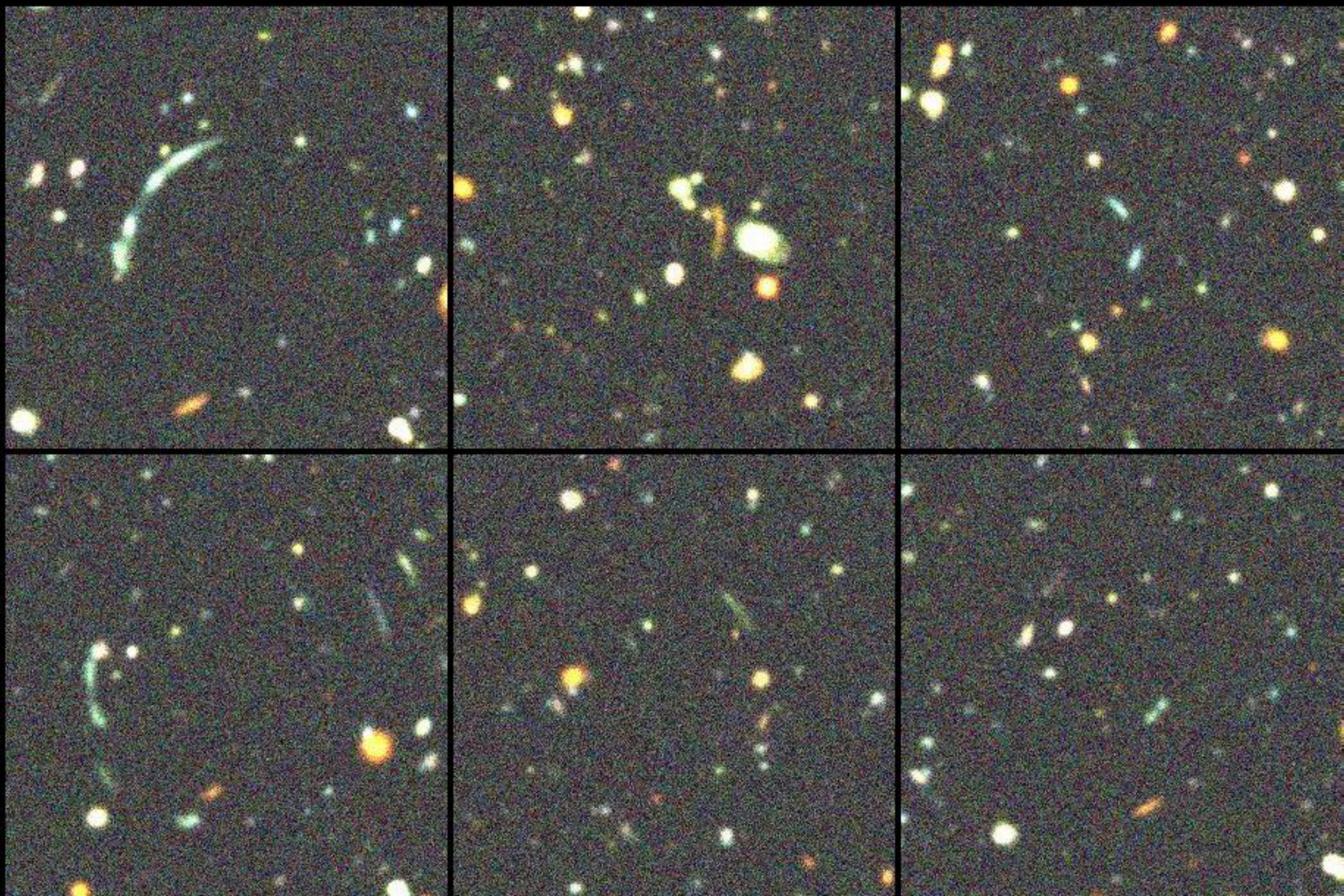}
\caption{``Observed'' examples of strong lensing, illustrating the classification process. Each image is 30"$\times$30", centered on the main halo. The two leftmost images are obvious instances of strong lensing and are identified as such with a score of 2 on the first pass; the central two images were classified as strong lensing on the next pass after further consideration. The rightmost pair of images were marked as ambiguous (a score of 1) on the first two passes, and only noted as strong lensing on the third pass after considering the higher-resolution noise-free raw input images and source object masks. In both of these cases that further examination (which can be thought of as analogous to the acquisition of follow-up data, and careful measurement) showed that the objects under consideration were indeed multiply-imaged and highly magnified. In the top-right image, the pair of quite elliptical blue sources near the center, with alignment suggestive of tangential shear, are indeed multiple images of the same source. In the bottom-right image, the dumbbell-shaped blue source to bottom right of center is a merging pair-image of one source, and a second and much fainter strongly lensed source can be seen on the opposite side of the lensing halo center.}
\label{fig:lensed_ims}
\end{figure*}

Two color images were made for each line of sight, one with the scaling adjusted to show the faintest structures, and the other with a stretch better sampling the full dynamic range of the input data. These images were then processed through a modified version of a visual search process that has been used previously \citep[e.g.,][]{Bayliss2011,Oguri2012}. In brief, each line of sight is quickly examined and given a score of 0, 1 or 2. A score of 0 means no evidence of any strong lensing; a score of 2 means obvious strong lensing, and a score of 1 means the proper classification is unclear at first glance. All objects given a score of 1 in this first classification "run" through the data were then re-examined and scored again in a second run after a more careful visual inspection as 0, 1 or 2. This two-fold examination allows for rapid broad cuts in the sample in order to process large numbers of images efficiently and focuses attention on ambiguous cases. In the above steps, both variants of color images were consulted as needed, though in practice the bulk of the classification comes primarily from the images with the broad dynamic range. Objects that were scored as ambiguous twice (in both the first run through all of the data, and in the second run through the data marked as ambiguous in the first pass) were then given a final classification in a final third run through the data, by consulting the raw input lensed data, absent PSF convolution or extra noise. Both single-halo and multi-halo lines of sight were considered in the same process, with the ordering of the input images randomized, and no indication given to the examiner of the identity of any input images. Scoring results were only unblinded after the entire scoring process was completed. Score distributions for the 20588 lines of sight considered are given in Table 1. Examples of image features with various non-zero scores are given in Figure \ref{fig:lensed_ims}. 

\vspace{0.25cm}
\begin{center}
\begin{tabular}{lrrr}
\hline
Score & Run 1 & Run 2 & Run 3 \\
 & (\# Images) &(\# Images) &(\# Images)\\
\hline
2&431 & 192 & 566 \\
1&1706 & 717 & 0 \\
0&18451 & 797 & 151 \\
All&20588 & 1706 & 717 \\
\hline
\end{tabular}
\tablecomments{Score distributions are for the first, second, and third classification runs. All objects marked as ambiguous (score = 1) are considered in the following classification run. }
\end{center}
\vspace{0.25cm}

This search process was designed to mock two types of strong lensing searches using optical imaging. The combined output of the first two classification runs gives the result to be expected from a careful visual analysis of only a set of survey images, absent any further data (e.g., spectroscopy or further imaging) or any detailed measurements that might further help quantify the likelihood of any features being due to strong lensing, such as colors or Gini coefficients \citep{florian2015a}. We note this, in what follows, as the ``survey-only'' sample. Adding the third classification run to that produces results more consistent with those expected from a survey in which extensive follow-up and analysis, beyond visual classification, is included. In what follows, this is referred to as the ``follow-up'' sample.

For all lines of sight with objects classified as strongly lensed, we then measured the magnification, multiplicity, length, and width of the lensed images, for up to 10 features in each image. The length and width are automatically measured from the ``observed'' images, using the lensed objects masks (if needed) to isolate flux from only the source of interest, by connecting all pixels in the observed image that correspond to that source and are more than $2\sigma$ above the sky noise. Together these can be used to compute the oft-used length-to-width ratio, $l/w$. The magnification is measured from the input data, absent PSF convolution and noise, as the ratio of the size of the source in the image plane to the size in the source plane. The multiplicity is similarly measured from the raw input images, just as the number of distinct pixel regions corresponding to the source of interest in the image plane.

\begin{figure*}[htp]
\centering
\includegraphics[width=0.49\textwidth]{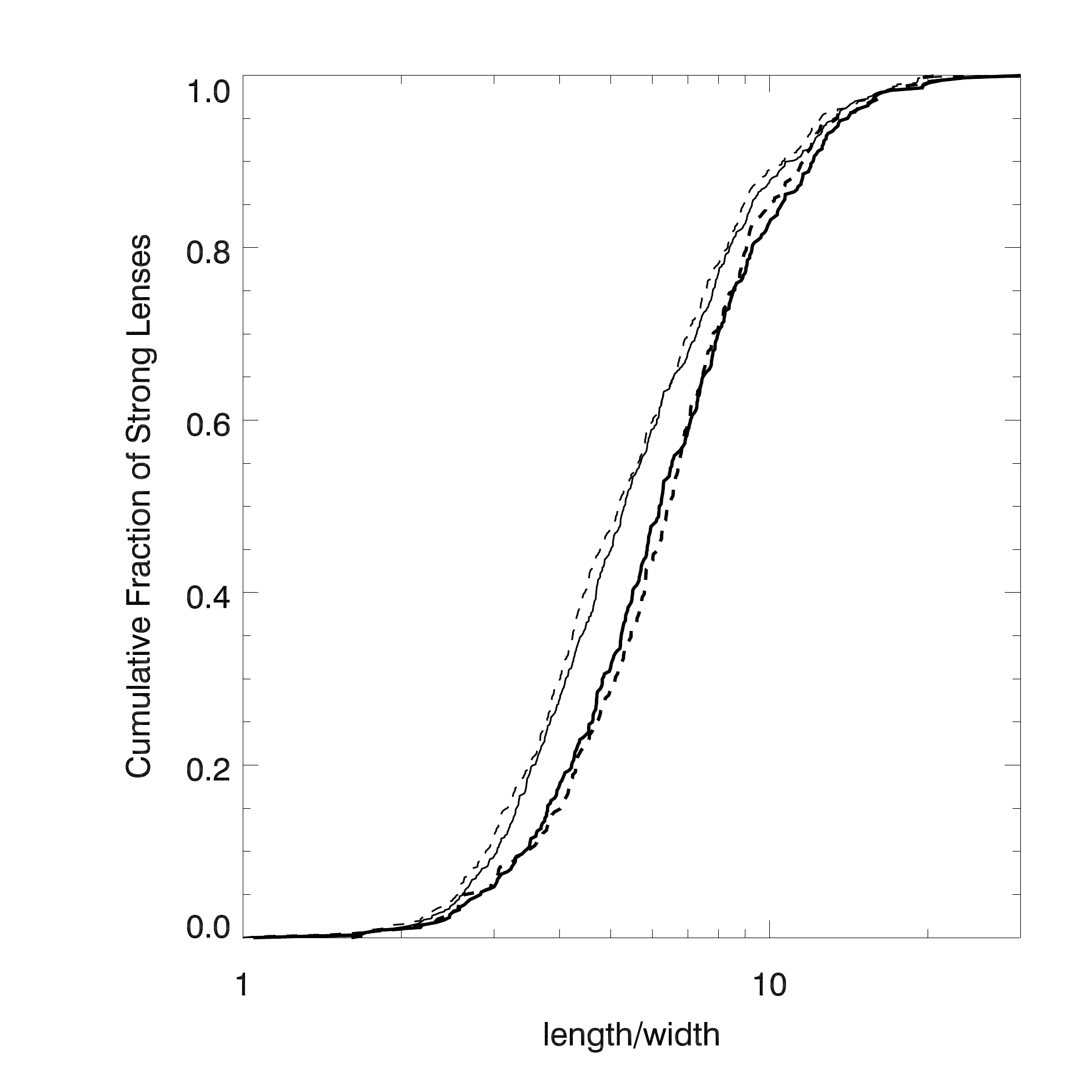}
\includegraphics[width=0.49\textwidth]{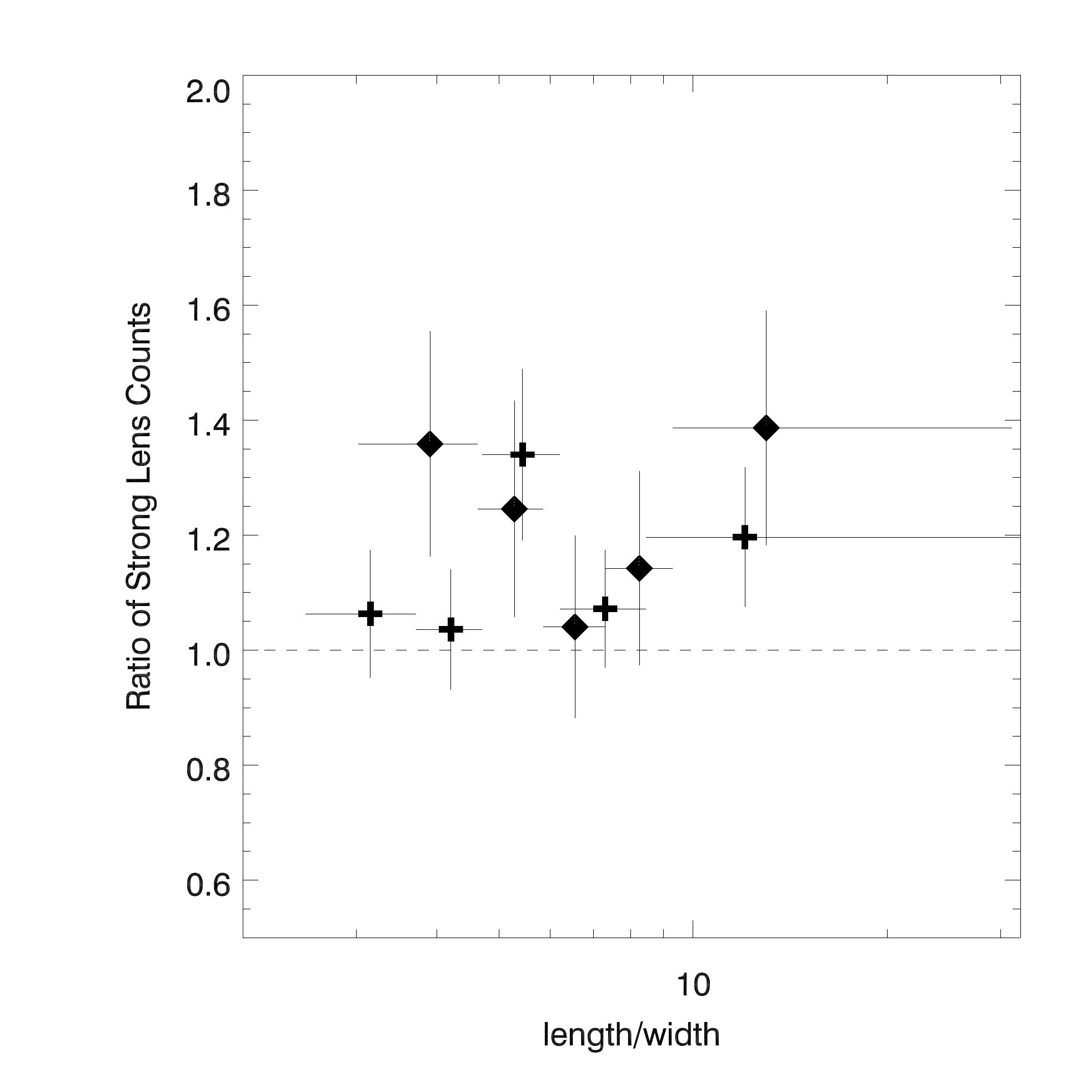}
\caption{Left Panel: The cumulative fractional distribution of $l/w$ values for the survey-only (thick lines) and follow-up samples (thin lines) for both the single-halo (dashed lines) and multi-halo cases (solid lines). Right Panel: The ratio of lens counts in the multi-halo to single-halo image simulations, in five $l/w$ bins with approximately equal total numbers of detected lenses, for the survey-only (diamonds) and follow-up (pluses) samples. Uncertainties are computed by bootstrapping the input halo list.}
\label{fig:lwsamp}
\end{figure*}

Before examining the lensing statistics in detail, it is worth noting that for 4493 of the halos considered, the single-halo and full-light cone input images are identical since the full light cone does not in some instances contain any additional halos within the angular range and mass limit considered. The observed images, which differ only in that each is one realization of the image noise, offer the opportunity to measure the repeatability and stability of the lensing search process since these two exceedingly similar images will have been present in the scoring process. In these lines of sight, each is considered twice; there are 149 strong lenses found in the aggregate survey-only sample; of these, 102 are found in both images, and 47 in only one of two images. The images identified only once are symmetrically distributed between the single- and multi-halo image sets, as expected.

The majority of the strong lenses identified only once are not entirely missed in the other image, however; of these 47 images, 36 were marked as ambiguous in the first two passes and considered in the follow-up sample, and of those 36, all but one were confirmed in that examination. The 11 remaining images found only once were all missed on the first pass examination of the other image. The bulk of the uncertainty built into the search process is thus classification uncertainty, rather than identification uncertainty.

\section{Results}
\label{sec:results}

In this section, we compare arc statistics with and without additional structures along the line of sight. For a given population of lensed sources, we consider the fractional change produced by using the full light cone, relative to the primary halo. We investigate strong lensing as a function of observed lensed source length-to-width ratio, and lens masses and redshifts. This is a focus driven by survey-scale observables; other properties (e.g., source magnifications, image multiplicity, and configurations, and so on) can be informative \citep[e.g.,][]{Oguri2001,Sand2005,Rozo2008} but are less readily measured across a large ensemble of strong lenses (particularly in ground-based data) and often require detailed modeling to be robustly extracted from a given dataset.

In analyzing the results below, it is also worth recognizing that the addition of other mass along the line of sight can not only create an instance of strong lensing -- it may also remove an instance. Additional halos along the line of sight will in general increase the length and apparent complexity of the lensing caustic structure (in the most extreme case, turning a sub-critical line of sight with no caustic structure into one that can strongly lens) in any given realization. However, it may also move the caustic in a way that it no longer intersects a given background source, removing that line of sight from the nominal strong lens catalog. The effect on lensing statistics of the full lightcone of halos is given by the sum across the entire sample of primary halos, and it is these statistics which are most relevant in comparison to actual samples of strong lenses. 

\subsection{The Effects of Secondary Halos as a Function of Lensed Image Length-to-Width Ratio}

The length-to-width ratio, $l/w$, is an observable property of lensed source images, and it is often used to isolate subsets of lensed images in both real data and simulations. As a measured property in real data, the $l/w$ is data dependent, since the width of typical lensed images is only poorly resolved in all but space-based imaging. As a reminder, the $l/w$ used here is taken from images convolved with a 0.7" PSF, and measured from all connected pixels of the largest image of a given source that are at least $2\sigma$ above the sky noise. Generally, these $l/w$ values will be smaller than one would measure from sharper images (by factors of at least several in comparison to Hubble Space Telescope data, for example), and smaller than one would measure from deeper images. 

Figure \ref{fig:lwsamp} shows the basic result for strong lens counts, for both the survey-only sample and the more complete follow-up sample. In this and the subsections which follow, each figure contains two presentations of the data: a cumulative fractional distribution showing the basic shape of the four potential subsamples (survey-only or follow-up, for either the single- or full multi-halo case) as well as a binned differential distribution of the multi- to single-halo count ratio, with uncertainties derived by bootstrapping the input halo list. As can be seen in Figure \ref{fig:lwsamp} from the turnover in counts at the lowest $l/w$ values (and as would otherwise be expected) the lens sample becomes incomplete at small source $l/w$ values, and so we choose to cut the samples at a lower limiting $l/w$ value, above which each is substantially complete. The chosen cuts are $l/w$=3.0 for the survey-only samples, and $l/w$=2.5 for the follow-up samples. To be clear, all samples of strong lenses are incomplete, in the sense that any particular survey will produce a sampling of the underlying population of all possible strong lenses, with the frequency of sampling given to first order by the background source density at the observing detection limits. The incompleteness at low $l/w$ is rather due to an inability to reliably distinguish strongly lensed images from a general galaxy in the light cone.

Incompleteness at the smallest values aside, the survey-only and follow-up samples differ significantly in $l/w$ distribution, as expected. The survey-only sample favors lensed sources with a larger $l/w$ since large $l/w$ images are more readily identified as due to strong lensing absent any further data (i.e., they are more morphologically distinct).
The mean $l/w$ in the survey-only sample is 6.3 versus 4.3 for the follow-up sample.

The ratio of multi- to single-halo lens counts is greater than one at all $l/w$ values in both the follow-up and survey-only samples. There is no significant evidence for a trend in this ratio with $l/w$, with at best a suggestion of a larger ratio in the largest 
$l/w$ bin. The measured mean ratio in the survey-only sample is $1.21\pm 0.07$, and in the follow-up sample it is $1.13\pm 0.05$; these differ insignificantly.

\subsection{The Effects of Secondary Halos versus Primary Lens Redshifts}
\label{sec:secvspl}

\begin{figure*}
\includegraphics[width=0.49\textwidth]{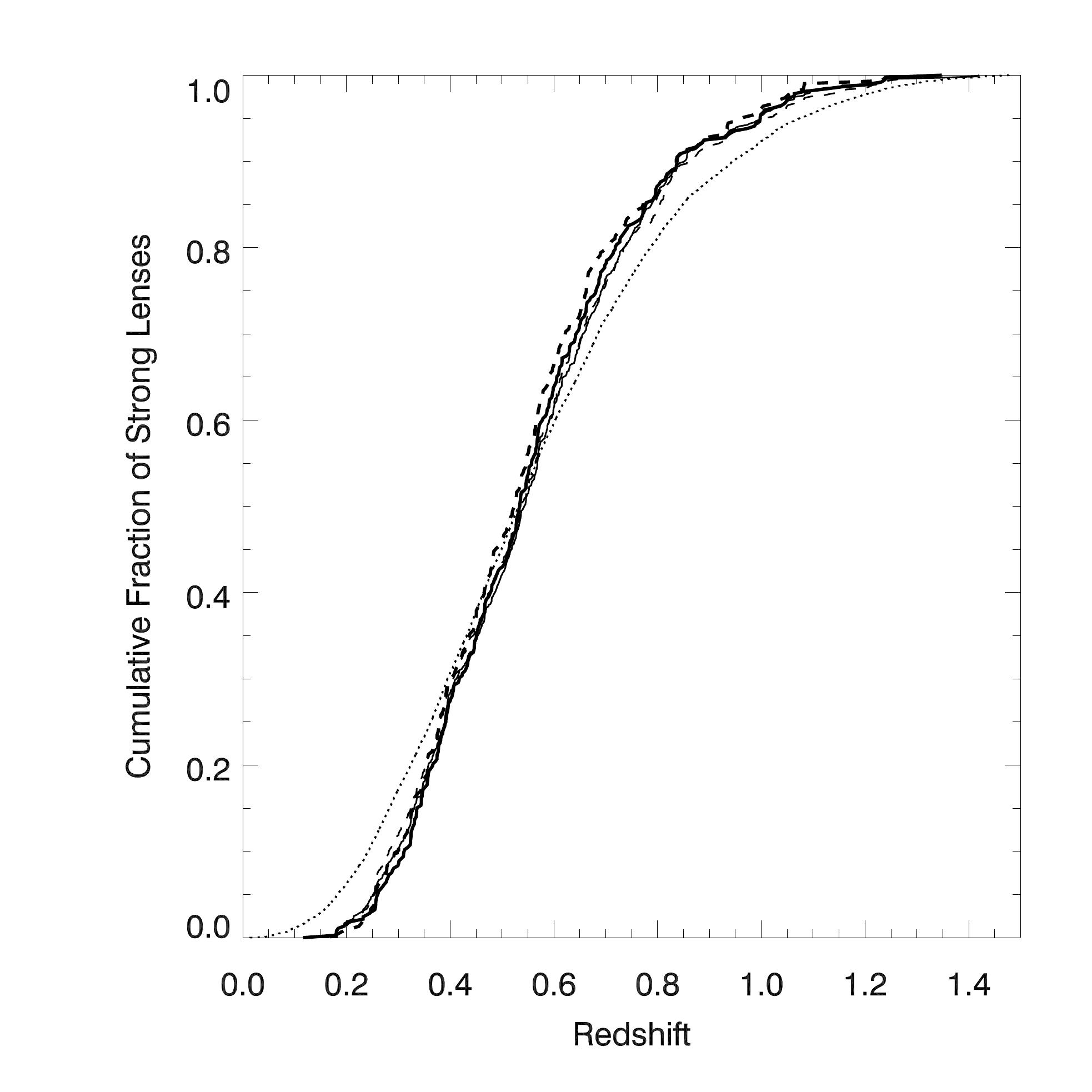}
\includegraphics[width=0.49\textwidth]{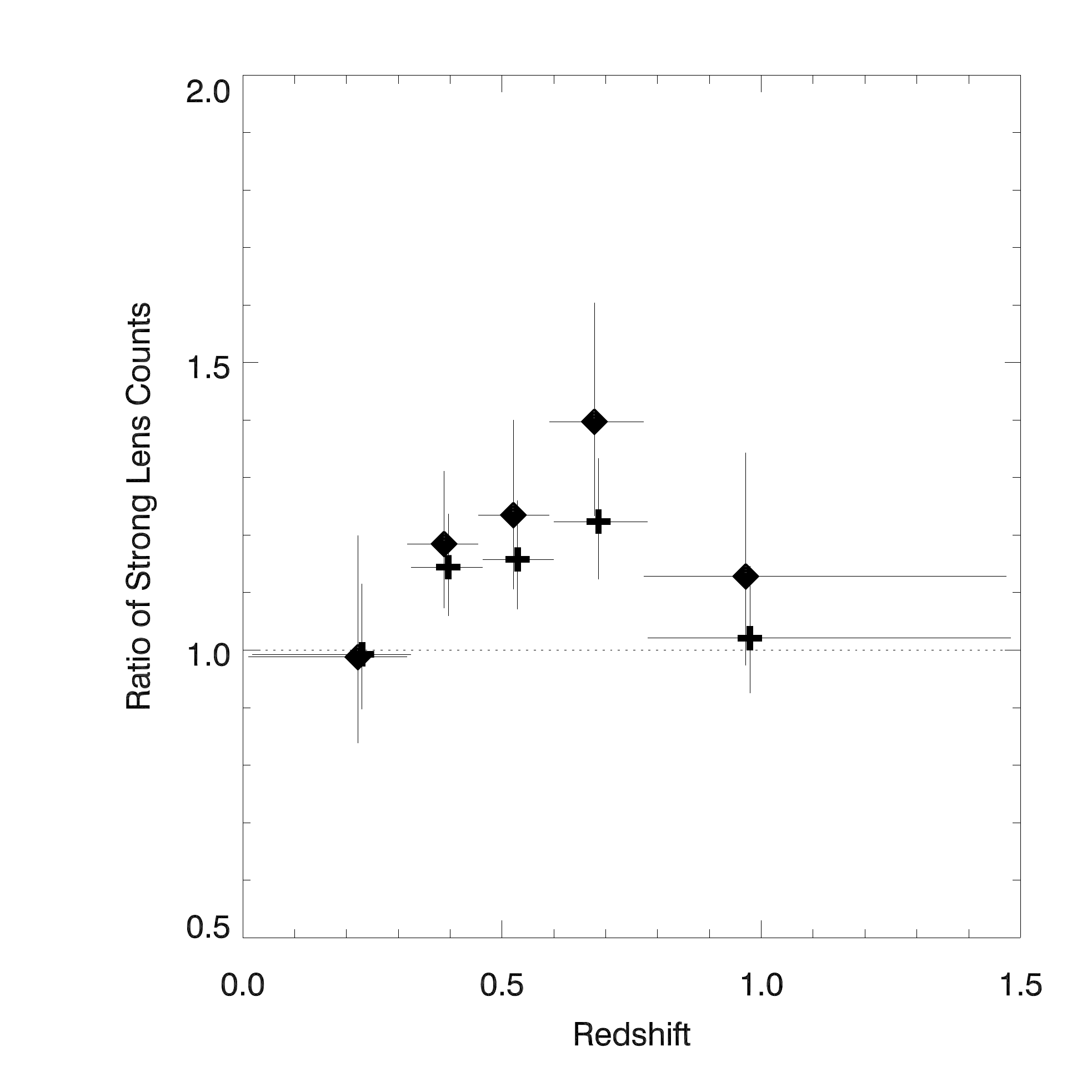}
\caption{Left Panel: The cumulative fractional distribution of redshift values for the various lens samples, denoted identically to Figure \ref{fig:lwsamp} with the addition of a dashed line to indicate the redshift distribution for the input halo catalog. Right Panel: The ratio of lens counts in the multi-halo to single-halo image simulations, in five redshift bins with approximately equal total numbers of detected lenses, also denoted identically to Figure \ref{fig:lwsamp}.}
\label{fig:redcdis}
\end{figure*}

The effects of secondary halos on the cross section of strong lensing as a function of primary lens redshift are shown in Figure~\ref{fig:redcdis}. Similar to Figure~\ref{fig:lwsamp}, the left panel of Figure~\ref{fig:redcdis} presents the cumulative fractional distribution of redshift values for the various lens samples. The curves have the same definition as those in Figure \ref{fig:lwsamp} with the addition of a dotted line to indicate the redshift distribution for the input halo catalog. As expected, the redshift distribution of the lens samples is significantly different from the input halo catalog. This effect is primarily geometric; the critical density for strong lensing scales as $D_s/(D_{ds}D_d)$, where $D_d$ is the observer to lens distance, $D_s$ is the observer to source distance, and $D_{ds}$ is the lens to source distance. For lenses at low redshift, the required critical density for strong lensing is thus quite large, and hence strong lenses are rarely seen. Strongly lensed sources are typically at $z\sim2$ \citep{Carrasco2017, Bayliss2011, Stark2013} at least in the wavelength and sensitivity regime explored here, and such a lens at high redshift has a small $D_{ds}$ and thus similarly the critical density for strong lensing is large and lensing uncommon. 

The right panel of Figure~\ref{fig:redcdis} shows the ratio of strong lensing counts -- multi- versus single-halo -- in five redshift bins. The influence of secondary halos on strong lensing counts peaks at about redshift $z\sim 0.6$, and the effect is marginally more significant in the survey-only dataset. In the lowest redshift bin, there is no observed effect, and the effect is marginal in the highest redshift quintile of the sample. 

The observed redshift distribution of the influence of secondary halos on strong lensing along the primary lines of sight is at least in part a reflection of the redshift distribution of the secondary halos, and the redshift threshold imposed on the secondary halo catalog. The secondary halos skew to higher redshifts than the primary halos, simply a reflection of the hierarchical nature of structure formation. Strong lensing happens close to critical curves, so to have a large effect, it is important that the distance from the secondary halos to the source rays which pass through the critical curves from the observer to the source is small.  When a secondary halo has the same redshift as the primary lens, the cross section for that secondary halo leading to significant influence will be a ribbon covering the critical curves of the primary halo. That is, the cross section for that secondary halo to significantly contribute to the strong lensing cross section depends on the extent of the critical curve of the primary halo. However, when the secondary halo has a different redshift from the primary lens, the cross section will be a ribbon covering \emph{rescaled} critical curves. The rescaled critical curves are the intersections between the rays (from the observer through the primary critical curves to the source) and the lens plane of the secondary halo. The rescaling relation is a linear function of the distance between the secondary halo and the primary lens. For a typical source, a cluster of fixed properties has the largest Einstein radius at intermediate redshift (peaking when $D_d=D_{ds}$), and couples best to secondary halos at the same redshift. Hence for a secondary halo population skewed to a higher redshift distribution than the primary halo population, the secondary halos are least effective in boosting the strong lensing at the lowest primary lens redshifts.

Also, it should be noted that the secondary halo catalogs used in these calculations are limited to $z<1.5$, and so the influence of secondary halos on the highest redshift primary halos may be artificially suppressed, and the results in Figure 3 should formally be considered as a lower limit particularly in the highest redshift bin.

Moreover, by considering the lensing effects as a function of halo redshift, there is also a potential interaction with the distribution of source properties (e.g., size, shape, and steepness of distribution with flux) with redshift, in that different halo subsets to some extent select different source populations. Quantifying this in detail would require detailed analysis of the source population, and is beyond the scope of this paper. 

\subsection{The Effects of Secondary Halos versus Primary Lens Masses}

\begin{figure*}
\includegraphics[width=0.49\textwidth]{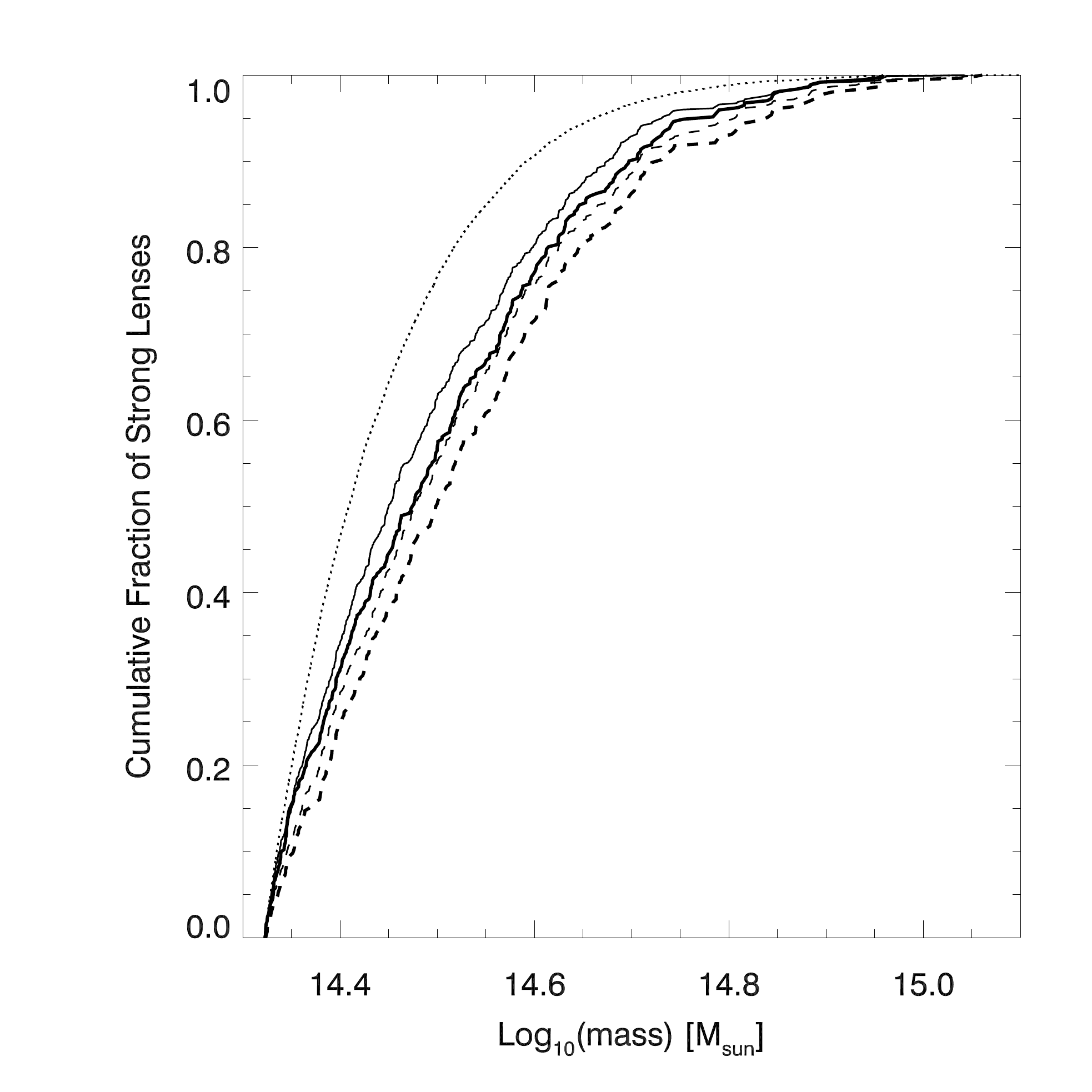}
\includegraphics[width=0.49\textwidth]{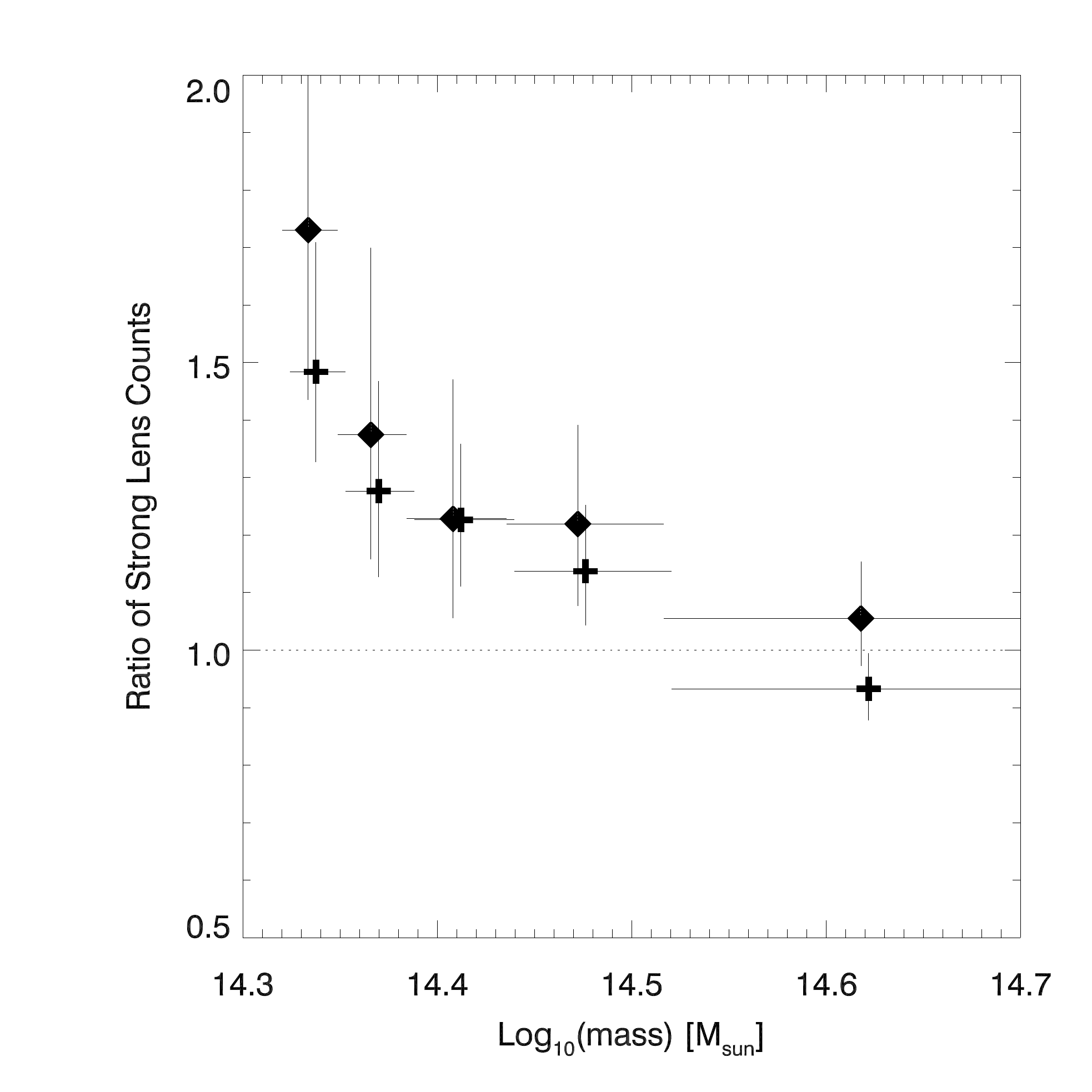}
\caption{Left Panel: The cumulative fractional distribution of strong lens mass values for the various lens samples, denoted identically to Figure \ref{fig:lwsamp}. The added dotted line is the cumulative mass fraction of primary halos. Right Panel: The ratio of lens counts in the multi-halo to single-halo image simulations, in five redshift bins with approximately equal total numbers of detected lenses, also denoted identically to Figure \ref{fig:lwsamp}.}
\label{fig:masscdis}
\end{figure*}

The influence of secondary halos on the strong lens samples, as a function of primary halo mass, is explored in Figure \ref{fig:masscdis}. The notation in Figure \ref{fig:masscdis} is identical to that of Figure \ref{fig:redcdis}. As anticipated, and shown in the left panel, more massive halos are more likely to act as strong lenses, and this remains true across both the multi- and single-halo simulations and the survey-only and follow-up samples. As before, the fractional increase due to secondary halos is considered in 5 bins spanning the range of primary halo masses considered with roughly equal numbers of halos per bin. In this instance, we see the strongest trend -- in the highest mass quintile there is no significant effect from secondary line-of-sight structures, but the influence of the line-of-sight increases sharply to produce a 50-80\% boost in lens counts in the lowest mass quintile.

This trend is due to the ratios of the total mass of the secondary halos to the primary lenses. For the most massive systems, secondary structures along the line of sight are typically insignificant relative to the primary halo. These primary halos are the most likely to be already critical to lensing; at lower masses, the addition of line-of-sight mass is much more likely to make the mass column critical to lensing. 

\section{Summary and Discussion}
\label{sec:sum}

For a sample of simulated massive halos more massive than $M_{500c}(\rho_{crit}) = 2.1\times 10^{14} \Msolh$, we find that line-of-sight mass structure increases the overall incidence rate of strong lensing by 13\%-21\%. This is for a search process using sub-arcsecond ground-based optical imaging with sensitivity as expected from a few minutes of integration time with an 8m class telescope, somewhat deeper but still relevant to the sensitivity of DES data. There is no significant correlation to lensed image length-to-width ratio. The boost in lens counts is maximal for primary halo redshifts at $z\sim0.6$, with no effect in the lowest quintile of the primary lens redshift distribution, and a lessened influence in the highest redshift systems. 

The effect of line of sight structures on strong lensing cross section is a steep function of primary cluster mass and runs counter to the impact of the primary halo itself. The net effect will be to flatten, to some extent, the dependence of overall lensing cross section on cluster mass (and reduce the overall sensitivity of arc statistics to details of the mass limit of a given real survey). However, it is clear that treatment of the mass limit and incompleteness at that limit for a given survey is needed to make detailed predictions for lensing statistics. The halo sample used here encompasses the published SPT-SZ cluster sample (median mass $\log_{10}(M_{500c}/\Msolh) \sim 14.54$) - that sample is of sufficiently high mass that it resides in the flattening tail of the distribution in Figure 4b. However, next-generation SZ cluster samples, such as that expected from the forthcoming SPTpol \citep{sptpol2012} data will have strong lensing considerably more affected by line of sight structure.

Previous results on the effect of line-of-sight mass on the detectability of strong lensing vary. \cite{Hennawi2007} found a 7\% increase in strong lensing, using a sample of 878 halos with ${\rm  M}_{vir}>10^{14} \Msolh$ taken from a simulation volume with a comoving box size of 320 $h^{-1}$ Mpc and a particle mass of $2.54\times10^9 \Msolh$, compared to 3000 $h^{-1}$ Mpc and $1.85\times 10^9 \Msolh$ for the simulation used here. \cite{Hennawi2007} see a trend with lensed image separation (a usable proxy for lens mass) in the same sense as found here -- with an increased effect to smaller image separations (i.e., generally lower lens masses). However, the overall amplitude of the effect seen in \cite{Hennawi2007} is significantly smaller than we observe; this difference is particularly significant given the mass trend seen in Figure \ref{fig:masscdis}, and the relative volumes and mass limits of the simulations used.  \cite{Puchwein2009} found an overall boost of 10-25\% from raytracing a set of cluster halos with the same mass limit as \cite{Hennawi2007}, taken from the Millennium Simulation \citep{Springel2005}, which has a comoving box size of 500 $h^{-1}$ Mpc and a particle mass of $8.6\times10^8 \Msolh$. Their quoted variation is a function of background source redshift; to extract the result for a particular set of observations requires information on the typical source redshifts. As noted in Section~\ref{sec:secvspl}, the typical source redshift for the observations simulated here is $z\sim2$, corresponding to the lower limit of the results from \cite{Puchwein2009}. 

One source of the difference may be cosmological; the simulation used in \cite{Hennawi2007}. used a power spectrum amplitude of $\sigma_8=0.95$, and the Millennium Simulation has $\sigma_8=0.90$, compared to the more observationally relevant value of  $\sigma_8=0.80$ for the Outer Rim simulation used here. At larger values of $\sigma_8$ massive cluster halos are more concentrated -- this demonstrably produces more occurrences of strong lensing \citep{Fedeli2008}, and this increased concentration may well make massive clusters less sensitive to the effects of secondary, albeit also more concentrated, mass structures along the line of sight. Qualitatively, the observations of \cite{Bayliss2014} favor a significant effect from secondary mass structures, as they see a significant enhancement in the projected line of sight mass structures in an observed sample of ten cluster strong lenses with extensive spectroscopy.

The analysis presented by \cite{Newman2013} argues that the total density profile of the central galaxy and the dark matter halo in massive clusters is described by an NFW profile \citep{1996ApJ...462..563N} consistent with the predictions of gravitational dynamics. \cite{Killedar2012a} find similar lensing cross sections between n-body and full hydrodynamical simulations of massive clusters. These result imply that baryonic effects do not significantly change the form of the density profile in massive cluster-scale halos. A gravity-only N-body simulation as used here is therefore appropriate for modeling strong lensing by massive clusters. At much lower masses -- that of galaxy-scale lensing, baryonic matter directly contributes significantly to the lensing cross section; between these two extremes, baryonic drag effects may couple the mass structure of light and dark matter, with varying degrees of predicted efficacy \citep{Puchwein2005, Rozo2008} in modifying strong lensing signals, depending in great part on the relative balance between heating and cooling in cluster cores \citep{Mead2010}. Exploration of these effects, from the observational perspective enabled by the PICS pipeline, including direct application to large hydrodynamical simulations, will appear in future work. However, it is worth noting here that effects such as baryonic drag or AGN heating that modify the central density of massive halos will also influence the degree to which strong lensing by said halos is influenced by other mass along the line of sight.

Finally, we note that lensing effects due to baryonic matter in the individual galaxies within more massive systems cannot ultimately be ignored, because gravitationally lensed arcs which are produced by galaxies are occasionally observed along the lines of sight near more massive lenses \citep{Halkola2006, Sand2008, Newman2009}. The complete and detailed comparison of observed strong lensing in a real cluster sample to that predicted from simulations must also consider this, with careful accounting not just for the existence but also cause of strongly lensed features seen. Differences between measured effects for line of sight projection in otherwise identical simulations \citep{Wambsganss2005, Hennawi2007} have been attributed to differences in the treatment of lensing signatures at small scales corresponding to galaxy-scale lensing \citep{Hennawi2007}. Though we have not attempted to explicitly separate galaxy-scale lensing from other signatures here, we note that the $l/w$ cut imposed on the analysis in part from the identification of lensing in simulated ground-based data, to a great extent precludes the inclusion of a significant number of true galaxy scale lenses. This is also indicated by the similarity of the single- and multi-halo lensed source $l/w$ distributions shown in Figure~\ref{fig:lwsamp}. 

We have adopted the analytic NFW profile to model the mass distribution of secondary halos on the line of sight because of the limits of our N-body simulation and density estimator. Shot noise will lead to more uncertainties if we insist on estimating the surface density maps of halos with particles when the numbers of particles are less than $10^4$. Observations of galaxy scale strong lenses demonstrate that the density profile is isothermal \citep[e.g.,][]{Auger2010} but that it evolves to NFW with increasing mass \citep{Shu2008}. Thus at the smallest mass scales of the secondary halo distribution we consider, we could be underestimating their lensing contribution. Similarly, previous studies have shown that elliptical lenses yield larger strong lensing cross-sections (e.g. \citealt{Meneghetti2003b} and \citealt{Oguri2003}), and so the use of simple spherical NFW profiles, caused by the limited number of particles in small halos, may also under-predict the effects of secondary halos. Additionally, we neglected the lensing effects of mass filaments along the line of sight due to the computational expense and the requirements of storage for building a full lightcone of all particles, which can in principle contribute to the lensing cross section as mass sheets. In practice, the critical density for lensing is sufficiently larger than the projected density of such structures such that they are unimportant in comparison to the uncertainties of current lens samples. Similarly, we also imposed thresholds of mass and redshift on the secondary halos when building the full sky light-cone; exclusion of some fraction of the potential mass and redshift space for secondary structures could thus also lead to an underestimate of the strong lensing impact of those structures. Therefore, the effects of secondary halos on the lensing cross sections of primary halos quantified in this paper are formally lower limits. 

Advanced cosmological simulations now being completed will allow for the inclusion of effects of secondary halos with lower masses and at higher redshift, more realistic surface density maps of said halos beyond a simple analytic NFW profile, the contributions from filaments and baryonic matter, and so on. Of likely greater immediate importance will be efforts to compare predictions of strong lensing which can be computed now to emerging large samples of strong lensing across a range of lens mass scales, and to refine simulations based on such comparisons. 

Details and caveats aside, the fundamental result we find is that secondary halos along the line of sight toward massive halos produce a 13-21\% increase in instances of strong lensing in typical moderately-deep ground-based imaging. With samples of lenses now in the hundreds, secondary mass along the line of sight must be included in future calculations of expected lensing samples, since the statistical uncertainty of emerging samples is sufficiently small to be sensitive to it. Given the trends observed with primary halo mass and redshift, line of sight mass can likely be ignored only for massive and low redshift halos; for example, the recent demonstration that the lensing cross-section of a small mass-selected cluster sample, mostly at low redshift, is consistent with theoretical expectations \citep{Xu2016} is likely not significantly compromised by secondary line of sight structures. This observed pattern of influence, with intermediate and higher redshift and lower mass primary halos showing the most effect, is suggestively consistent with the observed trends in halo concentrations in strong lensing galaxy clusters, in which low redshift and massive clusters are consistent with concentration  expectations for individual halos \citep{Merten2015}, but lower mass and higher redshift systems often are not \citep{Gralla2011, Oguri2012, Gonzalez2012, Foex2014}.

\acknowledgments{This research was funded in part by the Strategic Collaborative Initiative administered by the University of Chicago's Office of the Vice President for Research and for National Laboratories. Argonne National Laboratory's work was supported under the U.S. Department of Energy contract DE-AC02-06CH11357. This research used resources of the Argonne Leadership Computing Facility, which is supported by DOE/SC under contract DE-AC02-06CH11357. NL is also supported by a UK Science and Technology Facilities Council research grant. The work of HC was supported by the National Science Foundation Graduate Research Fellowship Program under Grants No. DGE-1144082 and DGE-1746045. MF's research was supported by an appointment to the NASA Postdoctoral Program at the NASA Goddard Space Flight Center, administered by Universities Space Research Association through a contract with NASA. This work was also supported in part by the Kavli Institute for Cosmological Physics at the University of Chicago through grant NSF PHY-1125897 and an endowment from the Kavli Foundation and its founder Fred Kavli. 
}

\bibliographystyle{apj}
\bibliography{./ms.bib}
\end{document}